# Estimating heterogeneous treatment effects from randomised trials: a comparison of the risk modelling and treatment effect modelling approaches

Frederik Luca Philipona[a,b], Marta Mainetti[b,c], Orestis Efthimiou[a,c,†], Georgia Salanti[a,†]

[a] *Institute of Social and Preventive Medicine (ISPM), University of Bern, Bern, Switzerland*

[b] *Graduate School for Health Sciences, University of Bern, Bern, Switzerland*

[c] *Institute of Primary Health Care (BIHAM), University of Bern, Bern, Switzerland*

[†] *co-last authors*

Corresponding author: Frederik Luca Philipona, frederik.philipona@unibe.ch.

Funding: Swiss National Science Foundation, grant number: 10002053

# Abstract

Introduction

Risk modelling (RM) and treatment effect modelling (EM) are two approaches to build models to estimate heterogeneous treatment effects from randomized control trials (RCT). RM is a two-stage approach; it estimates a predicted baseline risk at the first stage and then includes it as the only effect modifier in a regression model in the second stage. EM is a full treatment interaction multivariable model. Both approaches have theoretical advantages and limitations, but a thorough comparison including a simulation study is missing.

Methods

In a theoretical review of the two approaches, we present their underlying assumptions and theoretical advantages and disadvantages. Based on our theoretical review, we design simulation scenarios for RCTs with a dichotomous outcome and evaluate the performance of both approaches with respect to the root mean square error and the bias in the predicted risk difference.

Results

In the theoretical part we argue that baseline risk is a treatment effect modifier in many clinical situations. RM is a dimensionality reduction approach which, however, makes strong assumptions about the role of prognostic factors modifying the treatment effect. EM's greater flexibility is a possible advantage when the sample size is large.

In most simulated scenarios RM performs better than EM, even when the assumptions underlying RM are not fully met. The advantage of RM diminishes as sample size increases.

Conclusion

When choosing between RM and EM the available sample size and the plausibility of their underlying assumptions should be considered.

# 1 Introduction

Heterogeneous treatment effects is a term that describes treatment effects that change for subgroups of the participant population. Statistical models are developed to forecast the outcome for a particular participant with and without receiving a treatment, the expected treatment benefit, and guide treatment decision making[1]. Treatment effect modelling (EM) and risk modelling (RM) are two approaches to build such models using data from a randomised controlled trial (RCT). EM refers to a regression model where the treatment effect is modelled as a function of participant characteristics. RM is a two-stage approach where at the first stage a prognostic model is set up to estimate the risk of the outcome in the population (often called baseline risk); at the second stage the treatment effect is modelled as a function of the baseline risk of each participant.

Because EM is an intuitive way of building the prediction model, the literature comparing the two approaches has primarily focused on motivating the RM approach as an alternative2 3 4. RM aims primarily to address the problem of overfitting (as RCTs typically have small to moderate sample sizes); using a single interaction term between the predicted baseline risk and the treatment in stage two, RM is reducing the dimensionality of the model. Kent et al.1 note that baseline risk is a mathematical determinant of all treatment effect measures for dichotomous outcomes (risk difference (RD), relative risk reduction (RR) or odds ratio (OR)). Glasziou and Irwig [5] argue that for many medical interventions and outcomes studied in RCTs, baseline risk is an important treatment effect modifier. Indeed, a re-analysis of 32 large clinical trials showed substantial heterogeneity of treatment effects among participants with different predicted baseline risks4 for the outcome. The main argument for the use of EM in the literature is its flexibility. The PATH statement [3] argues that EM holds promise in larger datasets where shrinkage might be of less importance. Empirical comparison of the two approaches is limited. Van Klaveren et al.[6] compare RM and EM to predict the survival benefit of two treatments for participants with coronary artery disease and found large differences in resulting treatment recommendations. The EM model fit the data better than RM in terms of AIC. A simulation study was conducted[7] comparing RM and EM approaches. However, the RM approach used was substantially different from the two-stage approach commonly considered and described above: it was a one-stage regression model including only prognostic factors and a main treatment effect. In another simulation study[8] different RM approaches were compared between them, but not against EM. We could not identify a simulation study that compares EM and RM as defined here.

In this article, we address this gap by revisiting the two approaches, outlining their underlying assumptions, theoretical properties, strengths, and limitations, and compare their performance in an extensive simulation study. To ensure a fair comparison between the EM and RM approaches, we propose a flexible data-generating mechanism that does not assume either model is correctly specified and allows the degree of violation of their underlying assumptions to be systematically varied.

## 2 Notation and definition of the two approaches

We study the two approaches in the context of a single RCT with a dichotomous outcome of interest and an experimental intervention and a reference intervention; for convenience we will assume the reference is placebo. For simplicity we assume that the placebo intervention has no effect, such that the risk for the event of a participant randomized to the placebo arm is the risk had they not undergone any intervention. We call this risk the baseline risk and denote it by $\phi_0$. We denote the risk for the event in the treatment arm $\phi_1$. The trial has $n$ participants. The treatment indicator is $t_i$ for participant $i$. It is $t_i = 0$, if participant $i$ is randomized to the placebo arm, or $t_i = 1$, if participant $i$ is randomized to the treatment arm. The outcome for participant $i$ is $y_i = 1$, if the participant had the event of interest and $y_i = 0$, if not. Participant $i$ has participant characteristics $\boldsymbol{x_i} \in \mathrm{R}^{p_b+p_f+p_m}$. These participant characteristics are composed of $p_f$ pure prognostic factors (denoted with $\boldsymbol{f_i}$), $p_m$ pure effect modifiers (denoted with $\boldsymbol{m_i}$) and $p_b$ characteristics that are both effect modifiers and prognostic factors (denoted with $\boldsymbol{b_i}$). For varying values of a pure prognostic factor, the logOR between two interventions stays constant, while the overall risk of the outcome changes. For varying values of an effect modifier, the treatment effect in terms of logOR between two interventions changes. A pure effect modifier changes the treatment effect in terms of logOR but is not associated with the outcome independent on the treatment; this means the baseline risk stays constant.

For simplicity the prediction models considered are all logistic regression-based and only linear terms for the covariates are in the models.

The ***EM approach*** consists of a multivariable logistic regression model with main effects for prognostic factor characteristics and the treatment indicator and interactions between treatment and effect modifiers:

$$logit(E[Y_i|\boldsymbol{x_i}, t_i]) = \alpha^{EM} + \beta^{EM} t_i + \boldsymbol{\gamma_b^{EM}} \boldsymbol{b_i} + \boldsymbol{\gamma_f^{EM}} \boldsymbol{f_i} + \boldsymbol{\delta_b^{EM}} \boldsymbol{b_i} t_i + \boldsymbol{\delta_m^{EM}} \boldsymbol{m_i} t_i. \tag{1}$$

Here, $\alpha^{EM}$ is an intercept, $\beta^{EM}$ the main treatment effect, and $\boldsymbol{\gamma_b^{EM}}, \boldsymbol{\gamma_f^{EM}}$ the prognostic coefficients of participant characteristics $\boldsymbol{b_i}$ and $\boldsymbol{f_i}$, and $\boldsymbol{\delta_b^{EM}}, \boldsymbol{\delta_m^{EM}}$ the effect modifying coefficients of $\boldsymbol{b_i}$ and $\boldsymbol{m_i}$. In our simulation study we use an EM approach that does not explicitly differentiate between the different types of participant characteristics. We made this choice to simulate situations where the type of a participant characteristic is not known. We instead use LASSO penalization to select the right types of participant characteristics to fit model (1) in the simulation study.

In the ***RM approach***, a first stage uses a prognostic model to predict the outcome in the absence of the treatment, termed baseline risk. In our setting, the estimated baseline risk $expit(h_i)$ is obtained from a logistic regression model including only prognosis coefficients $\boldsymbol{\gamma_b^{RM_1}}, \boldsymbol{\gamma_f^{RM_2}}$ for the variables $\boldsymbol{b_i}$ and $\boldsymbol{f_i}$:

$$logit(E[Y_i|\boldsymbol{x_i}, t_i = 0]) = \alpha^{RM_1} + \boldsymbol{\gamma_b^{RM_1}} \boldsymbol{b_i} + \boldsymbol{\gamma_f^{RM_1}} \boldsymbol{f_i} = h_i. \quad (2)$$

This model can be fit to data of the trial being analysed (called RM with internal stage one) or to data external to the trial (RM with external stage one). Here, we only consider RM with internal stage one. It was shown that estimation of the baseline risk using both trial arms (ignoring information on the received treatment) provides better estimates than using only the placebo arm[7 9]; we will adopt this when we refer to the internal stage one.

The second stage of RM is a logistic regression model with a main treatment effect $\beta^{RM_2}$, and the estimated baseline log-odds $h_i$ as the only prognostic factor and effect modifier with $\gamma^{RM_2}$, $\delta^{RM_2}$ the predictive and effect modification coefficients:

$$logit(E[Y_i|\boldsymbol{x_i}, t_i]) = \alpha^{RM_2} + \beta^{RM_2} t_i + \gamma^{RM_2} h_i + \delta^{RM_2} h_i t_i. \quad (3)$$

# 3 Theoretical comparison of RM and EM approaches

The main difference between RM and EM is how they treat the baseline risk. RM assumes it is the only treatment effect modifier and in the first stage estimates it separately, while EM accounts for the possibility of baseline risk modifying the treatment effect by including the interaction terms for some of the prognostic factors with the treatment. With an illustrative example we first show why it is a good idea to account for baseline risk as an effect modifier. Then we explain possible advantages and disadvantages of how the two approaches treat the baseline risk. Lastly in this section, we discuss the remaining differences between the two approaches.

## 3.1 Baseline risk as a treatment effect modifier

Assume we design a very large RCT to investigate the effect of an experimental treatment. The participants have baseline risks uniformly distributed between 1% and 99% for an adverse outcome. They are independently of their baseline risk assigned to the treatment arm (where they undergo the experimental intervention) and the placebo arm. We assume for simplicity that the placebo intervention has no effect, such that a participant's risk for the event in the placebo arm is exactly their baseline risk. We call the risk in the placebo arm $\phi_0$ and the risk in the treatment arm $\phi_1$. We assume, for the purpose of illustration, that the experimental intervention halves the risk of the adverse event for each participant. In panels A and C of figure 1 we show the risk in the treatment arm ($\phi_1$) against the baseline risk ($\phi_0$) in black, and in panels B and D we show the treatment effect as a RD ($\tau = \phi_0 - \phi_1$) against baseline risk also in black. The RD is linearly increasing; the RD is highest for high baseline risk participants. Lastly, we assume that we have a RM stage one model whose predicted baseline risk $expit(h_i)$ for each participant $i$ is exactly this participant's true baseline risk $\phi_{0,i}$. We now fit two models to our data to try to predict the true individualized treatment effects.

We first fit model (a): $logit(E[Y_i|h_i, t_i]) = os(h_i) + \beta t_i$, where $os(h_i)$ means that the baseline log-odds enter the model as an offset term. This model can correctly predict the risks in the placebo arm but makes large errors for the prediction of the risks in the treatment arm (panel A of figure 1). It predicts treatment effects as a RD that are largest for participants at medium baseline risk and very small for high baseline risk participants (panel B of figure 1).

The second model we fit is model (b): $logit(E[Y_i|h_i, t_i]) = os(h_i) + \beta t_i + \delta h_i t_i$. This model includes baseline log-odds as a treatment effect modifier. We can see in panel D of figure 1 that this model is much better at predicting the individualized treatment effects.

The logistic regression model (a) without baseline log-odds as a treatment effect modifier imposes a specific relationship between the baseline risk and the treatment effect. It predicts that the experimental intervention only weakly reduces the risk for the adverse event for high baseline risk participants, and it predicts that the intervention will most reduce the risk for the event for medium baseline risk participants. This fact won't change for different values of $\beta < 0$ in model (a), although the exact shape of the treatment effect would change. Adding pure effect modifiers, $\boldsymbol{m}$, to model (a) will allow it to predict different treatment effects for the same baseline risk but on average the predicted risk reduction for high baseline risk participants will still be very small and it will be largest for medium baseline risk participants.

Whenever the treatment effect of an intervention does not follow the relationship imposed by model (a) (we postulate that this is often the case for clinical interventions) the baseline log-odds are an effect modifier on the log-odds scale.

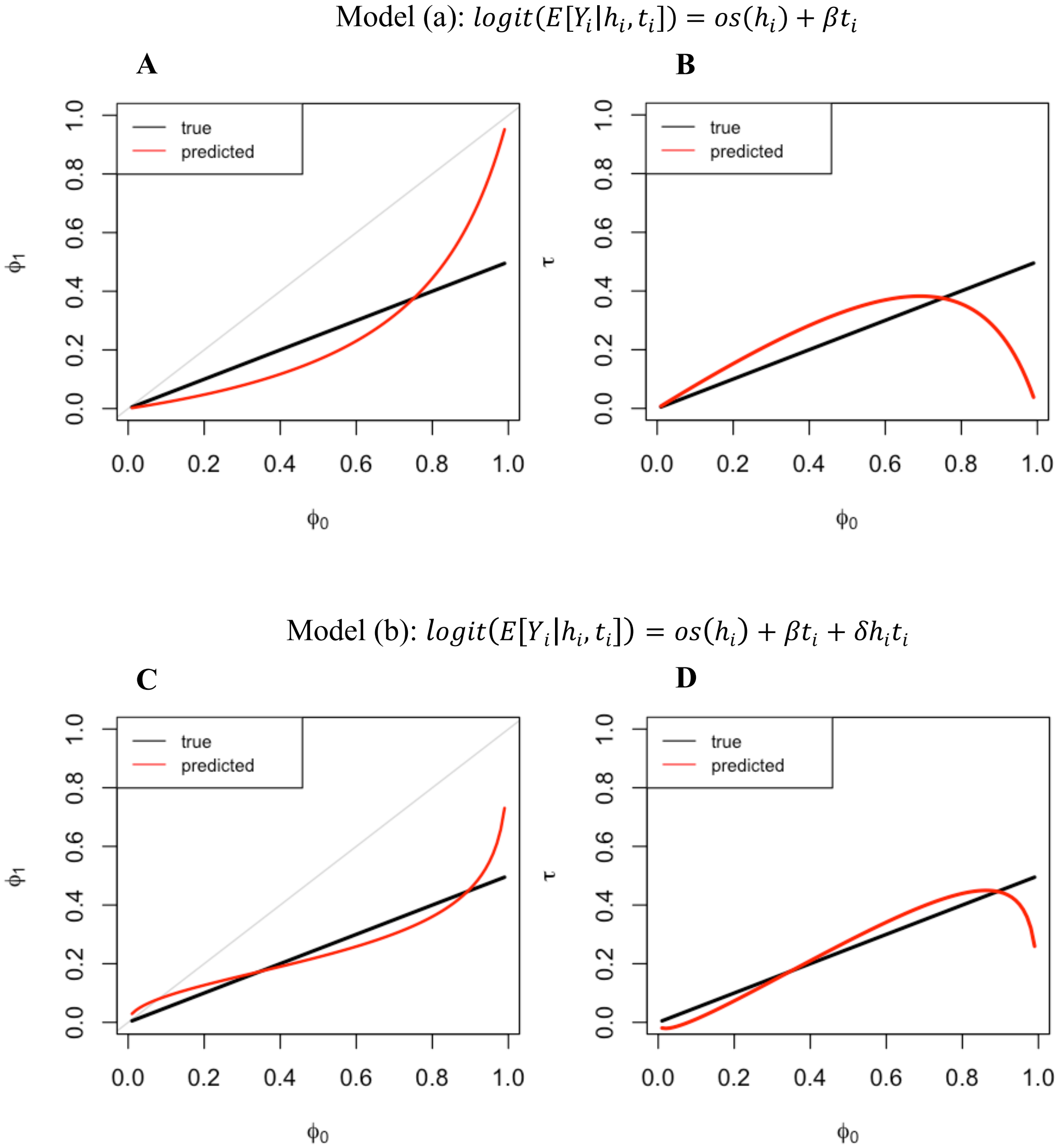

*Figure 1: True treatment risk and RD for the illustrative example against predicted risks and RD of two models. Panels A and C: risk in the treatment arm ($\phi_1$) against baseline risk ($\phi_0$). The diagonal lines are the risk in the placebo arm. Panels B and D: RD ($\tau = \phi_0 - \phi_1$) against baseline risk. Panels A and B show the predictions from model (a): $logit(E[Y_i|h_i, t_i]) = os(h_i) + \beta t_i$. Panels C and D show the predictions from model (b): $logit(E[Y_i|h_i, t_i]) = os(h_i) + \beta t_i + \delta h_i t_i$. The black lines are the true treatment arm risk and the true RD for the illustrative example of section 3.1*

### 3.2 Assumptions, advantages and disadvantages of the RM and EM approaches..

In the example situation presented in section. 3.1, the baseline log-odds are a treatment effect modifier on the log-odds scale. Both RM and EM can account for the baseline log-odds modifying the treatment effect.

The RM approach explicitly includes the interaction of treatment with the estimated baseline log-odds from the first stage model, which uses only prognostic factors $\boldsymbol{b}$ and $\boldsymbol{f}$. In the EM approach the prognostic factors are directly included in the model, with interaction with treatment (for participant characteristics $\boldsymbol{b}$) and without ($\boldsymbol{f}$). In the example in section 3.1 there exist no pure prognostic factors $\boldsymbol{f}$ and pure effect modifiers $\boldsymbol{m}$, because the baseline risk of each patient fully determines their treatment effect. In such a case any formula to predict individualized treatment effects obtained by RM can also be obtained by EM (see section 7.1 in the appendix). The two approaches then differ in how many coefficients they estimate: RM estimates $5 + p_b$ coefficients and EM estimates $2 + 2p_b$ coefficients. Then, if there are more than 2 participant characteristics, RM estimates fewer coefficients. This is a possible advantage of RM helping it protect against overfitting. However, this advantage comes at a cost that is associated with the assumptions underlying RM.

First, if pure effect modifiers $\boldsymbol{m}$ exist, their effect will not be captured in any of the RM stages. This is because, as they are not prognostic factors, their coefficient estimates in the first stage RM will be zero and they therefore won't contribute to $h_i$ in the second stage. Second, RM estimates only a single coefficient for $h_i$ at the second stage. From stage one, $h_i$ is the sum of what we call baseline log-odds contributions from $\boldsymbol{b}$ and $\boldsymbol{f}$ characteristics:

$$h_i = \alpha^{RM_1} + \gamma_{b,1}^{RM_1} b_{i,1} + \gamma_{b,2}^{RM_1} b_{i,2} + \cdots + \gamma_{b,p}^{RM_1} b_{i,p} + \gamma_{f,1}^{RM_1} f_{i,1} + \gamma_{f,2}^{RM_1} f_{i,2} + \cdots + \gamma_{f,r}^{RM_1} f_{i,r} \quad (4)$$

The coefficient $\delta^{RM_2}$ in the second stage of RM (equation (3)) assigns the same treatment interaction to all these baseline log-odds contributions. This means that the significance of a participant characteristic as an effect modifier is determined by its strength as a prognostic factor (the coefficients $\boldsymbol{\gamma_b^{RM1}}$). Finally, pure prognostic factors $\boldsymbol{f}$, will also contribute information to $h_i$ via the coefficients $\boldsymbol{\gamma_f^{RM1}}$ and hence they will modify the treatment effect. The EM approach does clearly separate the three types of characteristics – effect modifiers, prognostic factors, or both – and its greater flexibility is an advantage compared to RM. Under which circumstances, the flexible EM outperforms the parsimonious RM, is the subject of our simulation study that follows.

# 4 Comparison of RM and EM performance in simulations

We compare the performance of RM and EM in simulated data from an RCT where the treatment effect on a dichotomous outcome is a function of 6 participant characteristics. These characteristics can act as effect modifiers or prognostic factors (or both) in predicting the treatment effect and/or the outcome. Our simulation study is based on scenarios, which are combinations of four main scenarios and sub-scenarios. The main scenarios describe what types of participant characteristics are present, the sub-scenarios vary additional determinants of the performance of RM and EM such as sample size (section 4.1.2).

The treatment effect is, in the general case, not constant across RCT participants, but a function of their characteristics. We define the treatment effect function as difference between the risk of the outcome in the control arm $\phi_0(\boldsymbol{b}, \boldsymbol{f})$ minus the risk in the treatment arm $\phi_1(\boldsymbol{b}, \boldsymbol{f}, \boldsymbol{m})$:

$$\tau(\boldsymbol{b}, \boldsymbol{f}, \boldsymbol{m}) = \phi_0(\boldsymbol{b}, \boldsymbol{f}) - \phi_1(\boldsymbol{b}, \boldsymbol{f}, \boldsymbol{m}).$$

We assume that the risk in the control arm can be known by measuring only variables that are measured at baseline that act as prognostic factors. As in the rest of the article, we call the risk in the control arm hence baseline risk. The risk in the treatment arm can have a different form of dependence on the same prognostic factors and additionally depends on effect modifiers.

In the data generation mechanism, we assume a variety of clinically relevant treatment effect functions $\tau(\boldsymbol{b}, \boldsymbol{f}, \boldsymbol{m})$. The data are not generated using any of the RM or EM equations (1) and (3), to ensure a fair comparison. Instead, we assume scenarios where i) both the RM and EM models are misspecified and ii) the (stricter) assumptions of RM are violated to various degrees.

The simulation study was programmed in R version 4.5.1[10]. The packages glmnet[11], MASS[12], foreach[13], doParallel[14] and gplm[15] were used. The code can be found on GitHub: https://github.com/esm-ispm-unibe-ch-REPRODUCIBLE/RMvsEM.

## 4.1 Main- and sub-scenarios of the simulation study

Across main scenarios we vary the participant characteristics on which the treatment effect $\tau(\boldsymbol{b}, \boldsymbol{f}, \boldsymbol{m})$ depends. Across sub-scenarios we vary other determinants of relative performance of the RM and EM approaches, such as sample size. A "scenario" is the combination of a main- and a sub-scenario.

### *4.1.1 Main Scenarios*

**Main scenario BOTH:** There are 6 participant characteristics that are at the same time prognostic factors and effect modifiers (characteristics $\boldsymbol{b}$), the treatment effect is $\tau(\boldsymbol{b}, \boldsymbol{f}, \boldsymbol{m}) = \tau(\boldsymbol{b}) = \phi_0(\boldsymbol{b}) - \phi_1(\boldsymbol{b})$. In the RM approach all participant characteristics are at the same time prognostic factors and effect modifiers, like in this main scenario. When generating the data we, however, violate the RM assumptions by allowing the baseline log-odds contributions $(\gamma_{b,1} b_{i,1}, \dots, \gamma_{b,6} b_{i,6})^T$ to individually modify the treatment effect.

**Main scenario PF:** In this main scenario we only include 6 pure prognostic factors as participant characteristics. The treatment effect is $\tau(\boldsymbol{b}, \boldsymbol{f}, \boldsymbol{m}) = \tau(\boldsymbol{f}) = \phi_0(\boldsymbol{f}) - \phi_1(\boldsymbol{f})$. We set $\phi_0(\boldsymbol{f}) = expit(\boldsymbol{\gamma_f}^{\boldsymbol{T}} \boldsymbol{f})$ and $\phi_1(\boldsymbol{f}) = expit(\boldsymbol{\gamma_f}^{\boldsymbol{T}} \boldsymbol{f} + \beta_f)$, such that the treatment effect is a constant odds ratio of $exp(\beta_f)$. Only in main scenario PF, EM and RM are both correctly specified to predict the individualized treatment effects.

**Main scenario BOTH+PF:** There are 2 participant characteristics that are only prognostic factors (characteristics $\boldsymbol{f}$) and 4 characteristics that are both prognostic factors and effect modifiers (characteristics $\boldsymbol{b}$); the treatment effect is $\tau(\boldsymbol{b},\boldsymbol{f}) = \tau(\boldsymbol{b},\boldsymbol{f}) = \phi_0(\boldsymbol{b},\boldsymbol{f}) - \phi_1(\boldsymbol{b},\boldsymbol{f})$. The degree of misspecification of the RM approach for this main scenario is higher than under main scenario BOTH, because RM will use information from all prognostic factors to modify the treatment effect.

**Main scenario BOTH+PF+M:** The participant characteristics are: 1 pure effect modifier, 2 pure prognostic factors, and 3 characteristics that are both prognostic factors and effect modifiers; so, the treatment effect is $\tau(\boldsymbol{b},\boldsymbol{f},\boldsymbol{m}) = \phi_0(\boldsymbol{b},\boldsymbol{f}) - \phi_1(\boldsymbol{b},\boldsymbol{f},\boldsymbol{m})$. We consider this main scenario to be an unlikely case, as pure effect modifiers should be rare in practice. RM's first stage will not capture the pure effect modifier and therefore its effect modification will be omitted in the second stage. For this reason, this is the main scenario for which the RM approach is most misspecified to correctly predict the individualized treatment effects.

#### *4.1.2 Sub-scenarios*

Additionally, to the types of participant characteristics, we vary the following items in sub-scenarios.

**Relationship between treatment effect and baseline risk:** Whenever patient characteristics $\boldsymbol{b}$ are present in a main scenario (i.e in BOTH, BOTH+PF, BOTH+PF+M) a relationship between the treatment effect and baseline risk needs to be chosen. This relationship determines partly how the prognostic factors $\boldsymbol{b}$ modify the treatment effect. Importantly, in our data generation the relationship won't fully explain the individual treatment effects (see figure 2). Varying this relationship allows us to simulate data for a treatment that on average reduces the risk for high baseline risk patients strongly while only weakly reducing the risk for low baseline risk patients, and vice versa. We consider three different relationships between the treatment effect and baseline risk; these are shown in figure 2.

**Sample size**: Total sample size of 500, 2000 and 5000 participants per trial with 1:1 randomization. According to a recent review of clinical prediction models for heterogeneous treatment effects[16] sample sizes of 2000 to 5000 participants are common, while sample sizes of 500 participants are rare. However, most clinical trials have a smaller sample size than 500 participants[17] and the comparison of the performance of RM and EM for smaller sample sizes is therefore interesting.

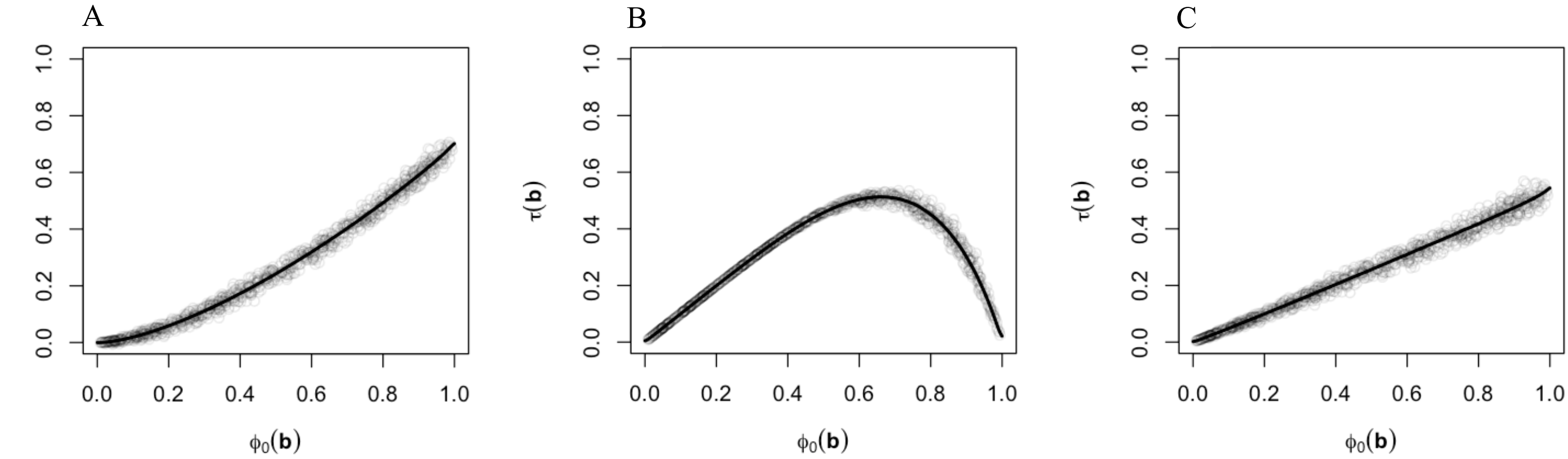

*Figure 2: The three relationships between the baseline risk and the treatment effect in the simulation study. Data was generated for main scenario BOTH. Black dots are generated treatment effects from one simulation run. The black lines represent the average treatment effects over all simulation runs. Panels A and C show a treatment with increasing effectiveness for increasing baseline risk. In panel C the increase is linear (similar to a risk ratio treatment effect), in panel A it is not. Panel B shows a treatment with decreasing effectiveness in baseline risk. For main scenarios BOTH+PF and BOTH+PF+M the simulated treatment effects are more widely dispersed around the same average effects and for main scenario PF the treatment effect is always a constant OR.*

**Distribution of baseline risk $\phi_0$**: We consider a realistic baseline risk distribution where most participants are at a low risk, and some are at a medium to high risk[4] (median = 35%, IQR = [20%, 60%], around 40% of the sample size are events on average), (approximately) uniform baseline risk (50% events on average), and rare event baseline risks (median = 5%, IQR = [1.8%, 14%], 11.5% average events). The baseline risk distributions for an example data generation run are shown in figure 3.

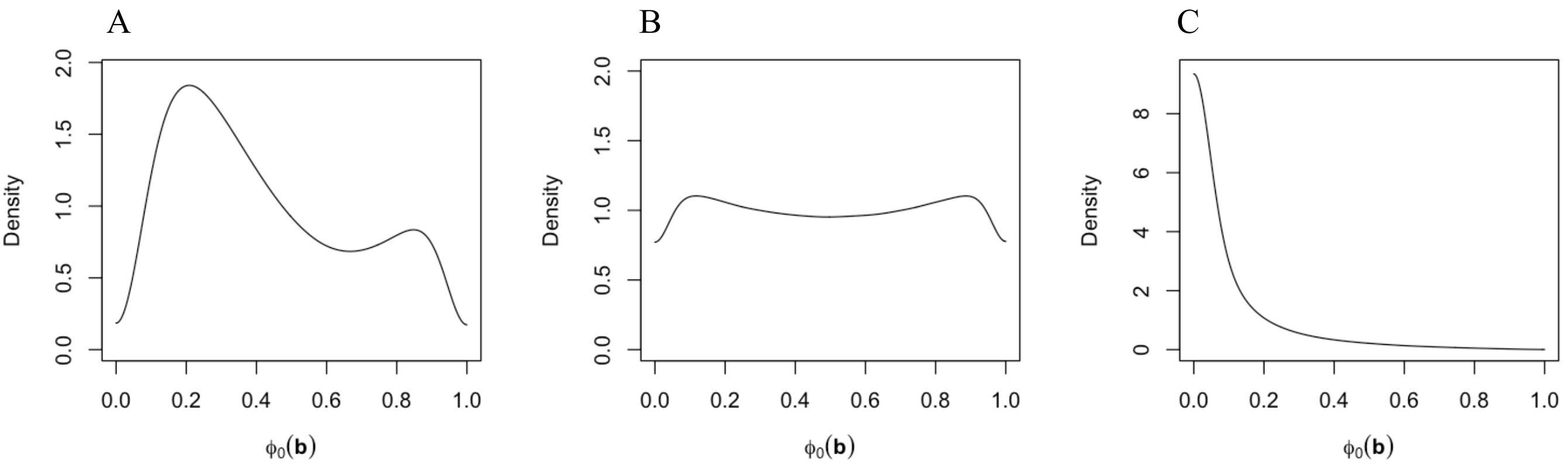

*Figure 3: Different baseline risk distributions from simulation runs of main scenario BOTH. Panel A: realistic baseline risk distribution, panel B: (approximately) uniform baseline risk distribution, panel C: rare outcomes risk distribution.*

**Effect modification for characteristics $\boldsymbol{b}$**: In the RM approach, each baseline log-odd contribution $\gamma_{b,1}b_{i,1}, \ldots, \gamma_{b,p}b_{i,p}$ is assumed to modify the treatment effect in the same way, as it is given the same regression coefficient. In our data generation we relax this assumption and set the difference in their effect modification to be small or large. Large differences in effect modification lead to a stronger misspecification of the RM approach for the prediction of the individualized treatment effects. Individualized treatment effects differ further from the average implied by the relationship of the baseline risk and the treatment effect (item above) when differences in effect modification are set to be large (see figure 4).

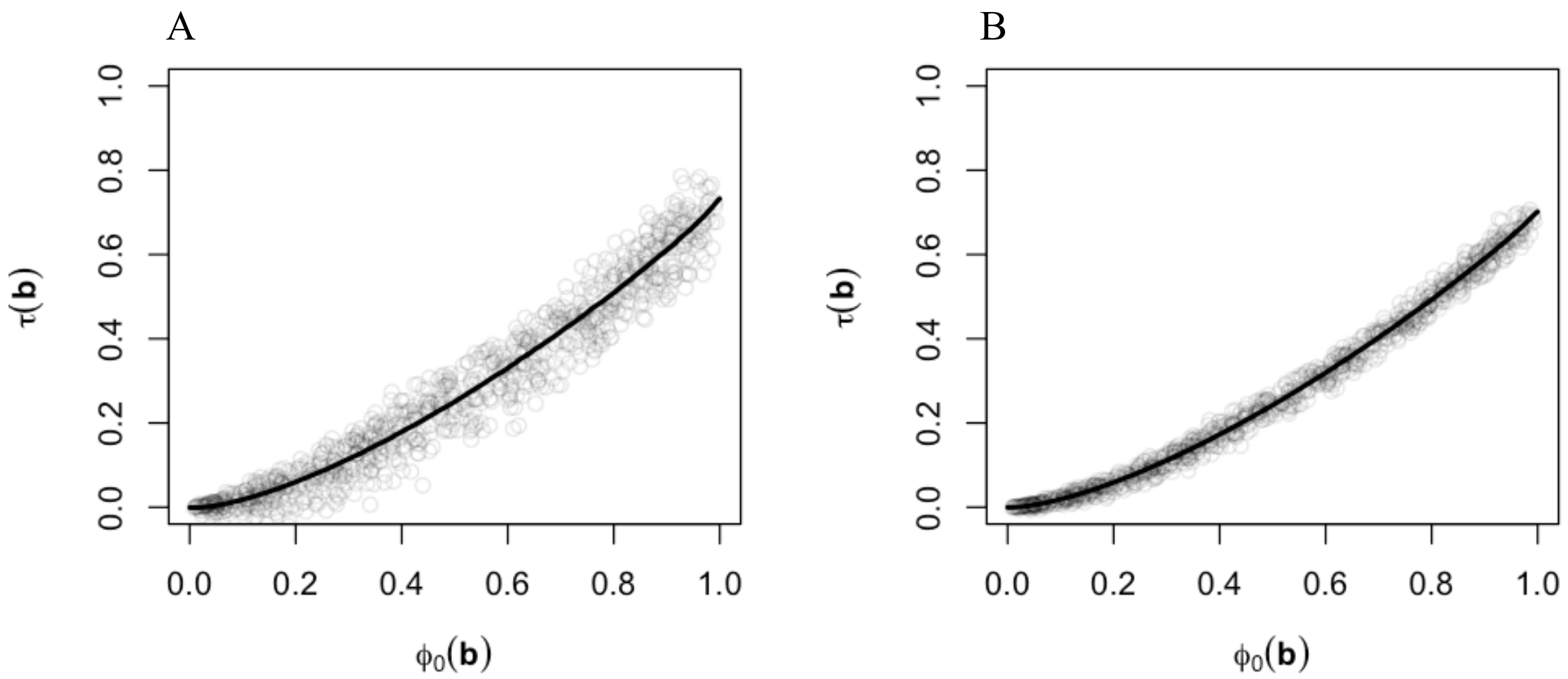

*Figure 4: Relaxing the assumption that the log-odds contributions of characteristics* $\boldsymbol{b}$ *have the same effect modification coefficient. Data was generated for main scenario BOTH for a treatment with increasing effectiveness for increasing baseline risk. Black dots are generated treatment effects from one simulation run. The black lines represent the average treatment effects over all simulation runs. The differences in effect modification of the characteristics* $\boldsymbol{b}$ *were set to large in panel A, and small in panel B. For main scenarios BOTH+PF and BOTH+PF+M the treatment effects are more widely dispersed about the line of the average treatment effects. For main scenario PF the treatment effect is always a constant OR.*

**Imperfect prognostic information**: RM performance depends a lot on the performance of the prognostic model in stage one, while capturing prognosis correctly is less important for EM. Here we exclude from the data seen by the analysis models the prognostic factor with the largest prognostic value in the data generation.

In table 1 we show all the combinations of main- and sub-scenarios we consider in our data simulation.

We generate the data samples in three steps: a) participant characteristics are drawn from normal distributions, b) risks for the event in the treatment and control arms are generated and c) events/non-events are generated for the treatment and control arms by drawing from Bernoulli distribution with the respective risks. The exact data generation formulae are shown in the appendix.

*Table 1: Sub-scenarios for the main scenarios*

| Main scenarios BOTH, BOTH+PF and BOTH+PF+M | | | | |
|---|---|---|---|---|
| Sub-scenario | $\boldsymbol{Baseline\ risk}$ distribution | Relationship $\boldsymbol{\phi_0}$ and treatment effect | Effect modification differences of $\boldsymbol{b}$ | Excluded $\boldsymbol{pf}$ |
| a) | Realistic | Effective for high and low baseline risk | Small | No |
| b) | Uniform | Effective for high and low baseline risk | Small | No |
| c) | Rare | Effective for high and low baseline risk | Small | No |
| d) | Realistic | Increasing effectiveness in $\phi_0$ | Small | No |
| e) | Realistic | Decreasing effectiveness in $\phi_0$ | Small | No |
| f) | Realistic | Effective for high and low baseline risk | Large | No |
| g) | Realistic | Effective for high and low baseline risk | Small | Yes |
| Main scenario PF | | | | |
| a) | Realistic | Treatment effect is a constant OR | Treatment effect is a constant OR | No |
| b) | Uniform | Treatment effect is a constant OR | Treatment effect is a constant OR | No |
| c) | Rare | Treatment effect is a constant OR | Treatment effect is a constant OR | No |
| d) | Realistic | Treatment effect is a constant OR | Treatment effect is a constant OR | Yes |

### 4.2 Analysis models and performance evaluation

The main analysis models were EM and RM. EM included all participant characteristics as covariates and their interactions with treatment and coefficients were estimated with LASSO penalization (all coefficient estimates were equally penalized). RM was fitted with an internal stage one, where data from both arms are used to fit the prognostic model using LASSO penalization, ignoring treatment assignment. The second stage of RM was estimated without penalization as it includes only one covariate. The LASSO penalization parameter, when applied, was chosen with ten-fold cross validation (minimizing the outcome prediction error, not the treatment effect error). The models were fitted using the R package glmnet[11].

The main performance metric we use is the out of sample root mean squared error (RMSE) of the predicted RD. The predicted RD for a participant is the difference between the predicted probability of the event without treatment minus the predicted probability of the event with treatment. Out of sample means that the prediction model coefficients are estimated in a trainings sample which is different than the test sample for which the performance metric is calculated. We also calculate the bias of the predicted RD of the two approaches.

### 4.3 Results of simulation study

Figure 5 shows the RMSEs of the treatment effect as a RD for all scenarios. The figure is split by sample size, as this was the biggest determinant of relative performance between the RM and EM approaches in the simulation study. The mean RMSEs are also shown in table 5 in the appendix.

RM outperforms EM in terms of mean RMSE in most scenarios tested for a sample size of 500 participants. In the other scenarios the performance is equal for the two approaches. For a sample size of 2000 participants RM still outperforms EM in most scenarios, but EM outperforms RM in some of the others. For a sample size of 5000 participants EM outperforms RM in about half the scenarios, RM outperforms EM in about a quarter of scenarios, and the performance in the other scenarios is equal. For large trials the performance of both approaches is very similar in most scenarios.

For main scenarios BOTH and PF, RM outperforms EM for almost all sample sizes and sub-scenarios. Main scenario BOTH includes only participant characteristics $\boldsymbol{b}$ that are at the same prognostic factors and effect modifiers. It is the main scenario which is most closely related to the assumptions underlying the RM approach. Main scenario PF includes only pure prognostic factors $\boldsymbol{f}$. RM achieves lower mean RMSE in this main scenario for most sample sizes and sub-scenarios because the LASSO penalization employed by EM does often not successfully set to zero all effect modifier coefficients.

Main scenario BOTH+PF includes participant characteristics $\boldsymbol{b}$ and $\boldsymbol{f}$, which corresponds to a softening of the RM assumptions. For this main scenario RM outperforms EM for sample sizes of 500 participants, for 2000 participants the better approach depends on the sub-scenario and for 5000 participants EM is better than RM in most cases.

When pure effect modifiers $\boldsymbol{m}$ are added in main scenario BOTH+PF+M, EM outperforms RM in most cases for sample sizes of 2000 and 5000 participants. For sample sizes of 500 participants, the performance of the two approaches is about equal. This main scenario is farthest away from the assumptions of the RM approach, and we consider the existence of pure effect modifiers to be very rare in practice.

Sub-scenario c), which tests rare events, favoured RM in our simulation study. Sub-scenario b), which tests uniform baseline risks, favoured EM. In both cases the change in the average number of events compared to the realistic baseline risks case changed the relative performance of the two approaches. A lower proportion of events favoured RM, and a higher proportion favoured EM. Sub-scenario f), which sets the difference in effect modification between the characteristics $\boldsymbol{b}$ to large, favored EM by a small amount. Both approaches were better able to predict treatment effects in situations with a treatment of decreasing effectiveness with increasing baseline risk (sub-scenario d)) compared to the case of a treatment with increasing effectiveness (sub-scenario e)), but this change had no impact on their relative performance.

For most scenarios in our simulation study both analysis methods achieved a very low overall bias of the treatment effect as a risk difference (see figure 6). Only for EM in case of a sample size of 500 participants the mean biases were slightly negative in all scenarios.

In figure 7 we show mean predicted RDs of the two analysis approaches against true simulated baseline risks for sub-scenario a) of all main scenarios with a sample size of 500 participants. Both analysis approaches tend to underpredict RDs for high baseline risk participants.

The biases for the highest quartile of baseline risk (table 7 in the appendix) were higher in magnitude for both analysis approaches and mostly negative. EM performed slightly worse in terms of bias for high baseline risk participants.

*Figure 5*: RMSE of the predicted treatment effect as a risk difference for the effect modelling (EM) and risk modelling (RM) approaches. The scenarios are presented in *Section 4*.

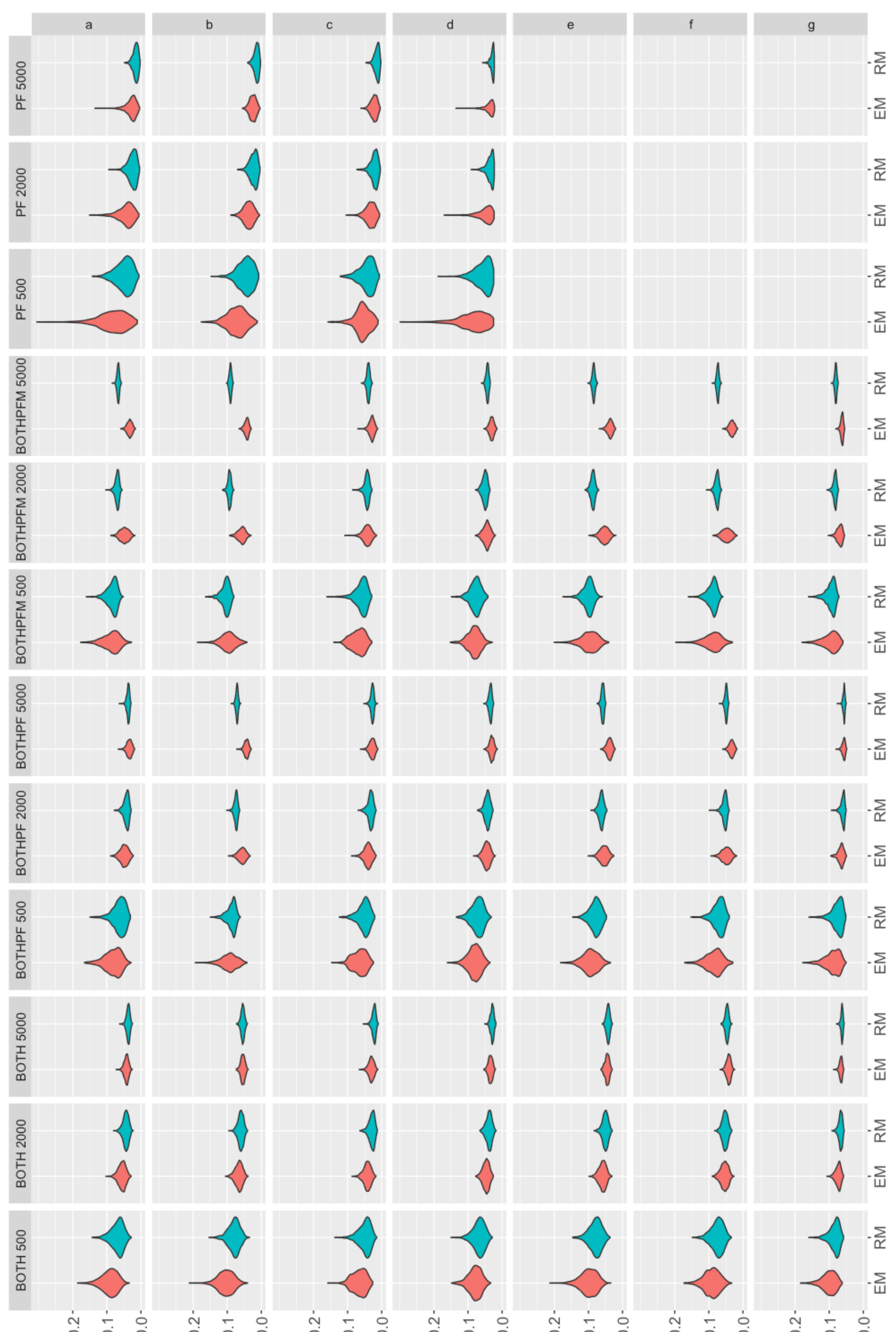

*Figure 6: Bias of* the predicted treatment effect as a risk difference *for the effect modelling (EM) and risk modelling (RM) approaches. The scenarios are presented in Section 4.*

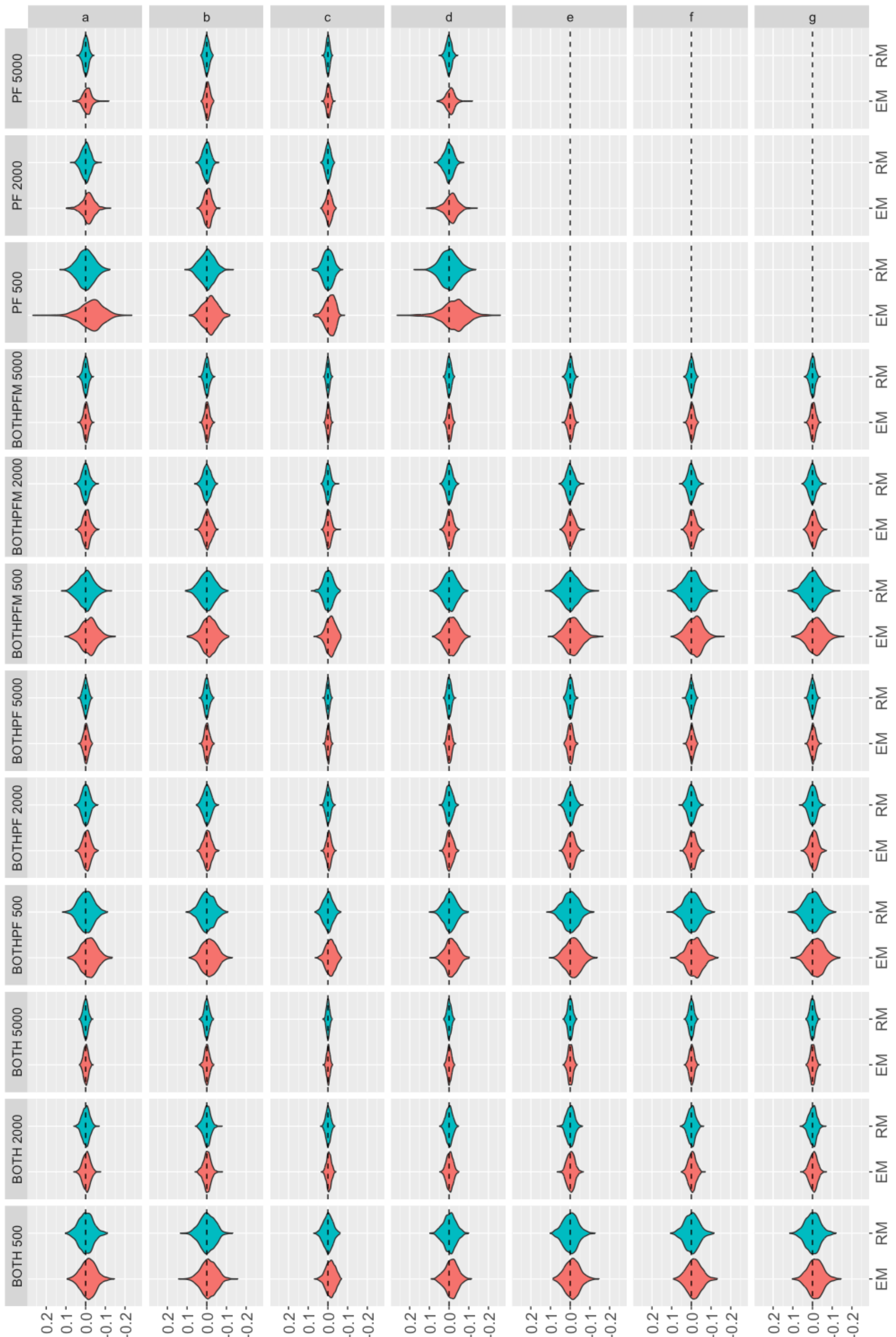

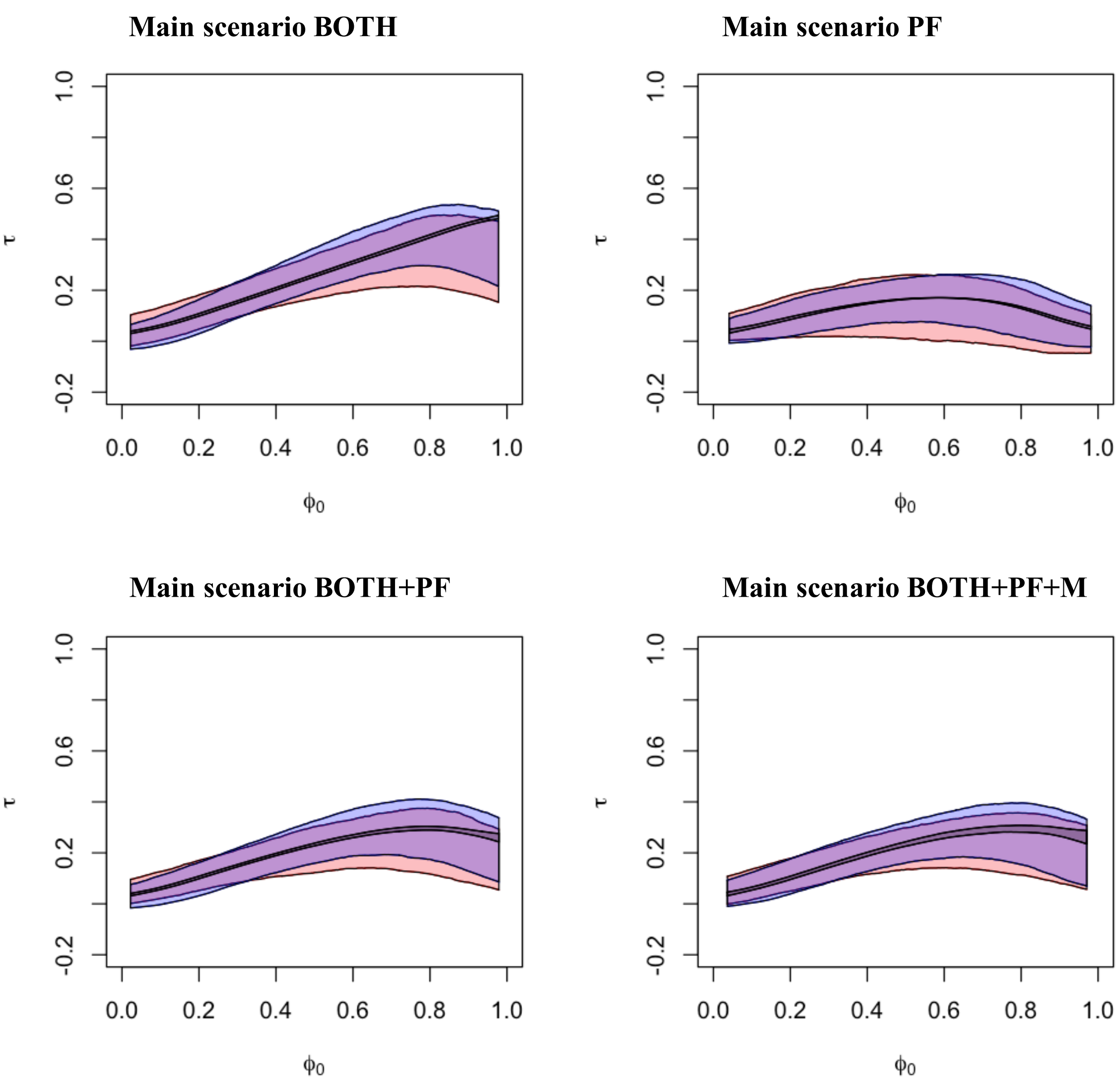

*Figure 7: Mean predicted risk difference τ (and 90% areas) against true baseline risk $\varphi_0$ for the effect modelling (EM, orange) and risk modelling (RM, blue) approaches for sub-scenario a) and sample size 500 participants. In the grey area lie 90% of the true simulated RDs. More details can be found in appendix section 1.4.*

## 5 Discussion

In this article we conducted a theoretical comparison and simulation study comparing RM and EM for the prediction of individualized treatment effects. The main difference between the two approaches is the role of the baseline risk. RM explicitly estimates baseline risk in the first stage and treats it as the sole treatment effect modifier in the second stage. In contrast, EM may include prognostic factors as effect modifiers, thereby allowing baseline risk to influence the treatment effect indirectly through these factors. We corroborated previous arguments in the literature that baseline risk is a treatment effect modifier in many clinical scenarios [3–5].

Our findings from an extensive simulation study demonstrate that sample size is the primary determinant of the relative performance of the RM and EM approaches. RM consistently outperformed EM in smaller samples (less than 2000 participants) and remained relatively robust when its underlying assumptions were violated. This advantage is likely attributable to the dimensionality reduction inherent to RM, which reduces model complexity and mitigates overfitting in settings with limited information. In contrast, the greater flexibility of EM becomes advantageous as sample size increases and more complex patterns of effect modification can be estimated reliably. RM performed particularly well when participant characteristics were prognostic factors or jointly prognostic factors and effect modifiers, whereas EM showed advantages in scenarios that included pure effect modifiers, especially in larger samples. Notably, both approaches exhibited low overall bias across most scenarios and showed comparable performance in large samples, suggesting that the choice between RM and EM should be guided primarily by the available sample size and the plausibility of their underlying assumptions.

Our study has some limitations. We considered only logistic regression-based RM and EM for dichotomous outcomes. Further simulation studies are needed to assess whether our findings extend to continuous and time-to-event outcomes. We also did not consider modern prediction modelling approaches, such as machine learning methods, which can be used to implement both RM and EM [18] [19] [20]. Similarly, we did not compare RM and EM with alternative approaches, such as meta-learners [21] [22]. These more elaborate methods may be particularly useful in settings with large sample sizes or observational data. Finally, we focused on the comparison of RM and EM in a single RCT. Although extensions of both approaches to meta-analysis and network meta-analysis have been proposed, we expect that the findings from the present simulation study will also apply in these settings [23] [24].

In summary, we provided the first direct comparison of the RM and EM approaches through an extensive simulation study. Our findings highlight a trade-off between the two methods: RM performs particularly well in settings with smaller sample sizes, whereas EM can offer advantages when larger samples allow more flexible modelling of treatment effect heterogeneity. For the prediction of individualized treatment effects, RM may therefore be preferred in small-sample settings, while the choice between approaches should consider both the available sample size and the plausibility of the underlying assumptions.

# 7 Appendix

## 7.1 Individualized treatment effect formula of RM obtained with EM

Any formula to calculate individualized treatment effects ($E[Y_i|x_i, t_i = 0] - E[Y_i|x_i, t_i = 1]$) where $E[Y_i|x_i, t_i]$ is calculated from the RM approach can also be obtained by the EM approach. To see this, we first combine the two stages of RM to get:

$$logit(E[Y_i|x_i, t_i]) = \alpha^{RM_2} + \beta^{RM_2} t_i + \gamma^{RM_2}\left(\alpha^{RM_1} + \gamma_b^{RM_1} b_i\right) + \delta^{RM_2} t_i \left(\alpha^{RM_1} + \gamma_b^{RM_1} b_i\right),$$

Where we assume there are only patient characteristics $b$. The EM model (formula (1)) looks like this:

$$logit(E[Y_i|\boldsymbol{x_i}, t_i]) = \alpha^{EM} + \beta^{EM} t_i + \boldsymbol{\gamma_b^{EM} b_i} + \boldsymbol{\delta_b^{EM} b_i} t_i.$$

With

$$\alpha^{EM} = \alpha^{RM_2} + \gamma^{RM_2}\alpha^{RM_1},$$

$$\beta^{EM} = \beta^{RM_2} + \delta^{RM_2}\alpha^{RM_1},$$

$$\gamma_{b,1}^{EM} = \gamma^{RM_2}\gamma_{b,1}^{RM_1}, \gamma_{b,2}^{EM} = \gamma^{RM_2}\gamma_{b,2}^{RM_1}, \dots, \gamma_{b,p_b}^{EM} = \gamma^{RM_2}\gamma_{b,p_b}^{RM_1} \text{ and}$$

$$\delta_{b,1}^{EM} = \delta^{RM_2}\gamma_{b,1}^{RM_1}, \delta_{b,2}^{EM} = \delta^{RM_2}\gamma_{b,2}^{RM_1}, \dots, \delta_{b,p_b}^{EM} = \delta^{RM_2}\gamma_{b,p_b}^{RM_1},$$

we get the same formula ($E[Y_i|\boldsymbol{x_i}, t_i = 0] - E[Y_i|\boldsymbol{x_i}, t_i = 1]$) from both approaches.

## 7.2 Additional information on data generation

We generate the data samples in three steps: a) participant characteristics are drawn from normal distributions, b) risks for the event in the treatment and control arms are generated and c) events/non-events are generated for the treatment and control arms by drawing from Bernoulli distribution with the respective risks.

### *7.2.1 Participant characteristics*

For the 5 (main scenario BOTH+PF+M) or 6 prognostic participant characteristics $\boldsymbol{b}$ and $\boldsymbol{f}$ we draw from a multivariate normal distribution $N(\boldsymbol{\mu}, \sigma\boldsymbol{R})$. The correlation matrix $\boldsymbol{R}$ has entries $\boldsymbol{R_{i,i}} = 1$ and $\boldsymbol{R_{i,j}} = 0.25^{|i-j|}$ if $i \neq j$. The mean vector $\boldsymbol{\mu}$ and $\sigma$ depend on the desired baseline risk distribution. We set $\sigma$ to 1 for uniform and rare baseline risks and to 0.25 for realistic baseline risks.

For uniform baseline risks the mean vector is $(0,0,0,0,0,0)^T$ and for rare outcomes it is $(-1,-1,-1,-1,-1,-1)^T$. For the realistic case a participant is first either classified as a high or low risk participant using a Bernoulli distribution with the probability of 0.25 to be in the high-risk group.

Then, for a high-risk participant the mean vector is $\left(1.5/(6\gamma_1), 1.5/(6\gamma_2), 1.5/(6\gamma_3), 1.5/(6\gamma_4), 1.5/(6\gamma_5), 1.5/(6\gamma_6)\right)^T$ and for a low-risk patient it is $\left(-1/(6\gamma_1), -1/(6\gamma_2), -1/(6\gamma_3), -1/(6\gamma_4), -1/(6\gamma_5), -1/(6\gamma_6)\right)^T$. In these formulas $\gamma_{\mathrm{j}}$ is the coefficient of the prognostic factor $j$.

After having drawn from the multivariate normal distributions two of the prognostic factors are dichotomized by assigning a value of 1 if the drawn value for that prognostic factor is greater than 0 and 0 otherwise.

### *7.2.2 Generation of control and treatment risks*

The formulae to generate the control and treatment arm risks in each main scenario are summarized in table 2. The exact coefficients in the formulae for each combination of main- and sub-scenario are shown in table 3. We now explain the risk generation formulae.

Consider first our main scenario BOTH, with only participant characteristics $\boldsymbol{b}$, which are at the same time prognostic factors and treatment effect modifiers. In this main scenario the risk of participant $i$ in the control arm is

$$\phi_0(\boldsymbol{b_i}) = expit(\gamma_b b_i),$$

where $\boldsymbol{\gamma_b b_i}$ are the baseline log-odds of participant $i$. If we would then set the risk for participant $i$ in the treatment arm to

$$\phi_1(\boldsymbol{b_i}) = expit(\boldsymbol{\gamma_b b_i}) \times 0.5$$

the treatment effect would be a constant risk ratio of 0.5 for all participants. This corresponds to the situation where a treatment is effective for high and low baseline risk participants. The characteristics $\boldsymbol{b}$ would all modify the treatment effect in the way determined by the constant risk ratio. The individual baseline log-odds would fully explain the individualized treatment effects, which corresponds to the assumptions of the RM approach.

Instead of the above, we generate the risk in the treatment arm like this:

$$\phi_1(\boldsymbol{b_i}) = expit\left(logit(expit(\boldsymbol{\gamma_b b_i}) \times 0.5) + \boldsymbol{\delta_b}\left(\boldsymbol{\gamma_b}^{\boldsymbol{T}} \circ \boldsymbol{b_i}\right)\right), \boldsymbol{\delta_b} \neq \boldsymbol{0}. \quad (4)$$

Here, $(\boldsymbol{\gamma_b} \circ \boldsymbol{b_i})$ denotes the vector of the entry-wise products of $\boldsymbol{\gamma_b}$ and $\boldsymbol{b_i}$. This is the vector of baseline log-odds contributions. With the vector $\boldsymbol{\delta_b}$ we let the baseline log-odds contributions individually modify the treatment effect. The individual baseline log-odds now no longer fully explain the treatment effect. To allow for different relationships between baseline risk and the treatment effect we write $expit(\boldsymbol{\gamma_b b_i}) \times \beta_b(\boldsymbol{\gamma_b b_i})$ instead of the term $expit(\boldsymbol{\gamma_b b_i}) \times 0.5$ in (4). For different choices of $\beta_b(\boldsymbol{\gamma_b b_i})$ we can simulate treatments that are, on average, highly effective for high baseline risk participants and weakly effective for low baseline risk patients, and vice-versa (see table 3).

In main scenario BOTH+PF there are participant characteristics $\boldsymbol{b}$ which are both prognostic factors and effect modifiers and participant characteristics $\boldsymbol{f}$ which are pure prognostic factors. We set the risk in the control arm in this main scenario to

$$\phi_0(\boldsymbol{b_i}, \boldsymbol{f_i}) = expit(\boldsymbol{\gamma_f f_i} + \boldsymbol{\gamma_b b_i}),$$

where $\boldsymbol{\gamma_f f_i} + \boldsymbol{\gamma_b b_i}$ are the baseline log-odds of participant $i$. The risk in the treatment arm for participant $i$ in main scenario BOTH+PF we set to

$$\phi_1(\boldsymbol{b_i}, \boldsymbol{f_i}) = expit\left(\boldsymbol{\gamma_f f_i} + logit\big(expit(\boldsymbol{\gamma_b b_i}) \times \beta_b(\boldsymbol{\gamma_b b_i})\big) + \boldsymbol{\delta_b}\big(\boldsymbol{\gamma_b}^{\boldsymbol{T}} \circ \boldsymbol{b_i}\big)\right), \boldsymbol{\delta_b} \neq \boldsymbol{0}.$$

This data generation model again includes a relationship between parts of the baseline risk and the treatment effect, but it explains less of the individualized treatment effects as in main scenario BOTH. For varying values of the participant characteristics $\boldsymbol{f}$ the treatment effect as a log-odds difference does not change.

In main scenario BOTH+PF+M we add additional pure effect modifiers $\boldsymbol{m}$. The risk for participant $i$ in the control arm is the same as in main scenario BOTH+PF because the additional pure effect modifiers don't change the baseline risks. The risk for participant $i$ in the treatment group in main scenario BOTH+PF+M is

$$\phi_1(\boldsymbol{b_i}, \boldsymbol{f_i}, \boldsymbol{m_i}) = expit\big(\boldsymbol{\gamma_f f_i} + logit\big(expit(\boldsymbol{\gamma_b b_i}) \times \beta_b(\boldsymbol{\gamma_b b_i})\big) + \boldsymbol{\delta_b}\big(\boldsymbol{\gamma_b}^{\boldsymbol{T}} \circ \boldsymbol{b_i}\big) + \boldsymbol{\delta_m m_i}\big), \boldsymbol{\delta_b} \neq \boldsymbol{0}, \boldsymbol{\delta_m} \neq \boldsymbol{0}.$$

Finally, we have main scenario PF which has only patient characteristics that are pure prognostic factors. The risk of participant $i$ in the control arm is

$$\phi_0(\boldsymbol{f_i}) = expit(\boldsymbol{\gamma_f f_i}),$$

and the risk of participant $i$ in the treatment arm is

$$\phi_1(\boldsymbol{f_i}) = expit(\boldsymbol{\gamma_f f_i} + \beta_f).$$

The treatment effect in main scenario PF is a constant log-odds difference $\beta_f$.

*Table 2: Formulae to generate the risks in the control and treatment arms.*

| Main scenario | $\phi_0(\boldsymbol{b_i}, \boldsymbol{f_i})$ | $\phi_1(\boldsymbol{b_i}, \boldsymbol{f_i}, \boldsymbol{m_i})$ |
|---|---|---|
| BOTH | $expit(\boldsymbol{\gamma_b b_i})$ | $expit\left(logit\big(expit(\boldsymbol{\gamma_b b_i}) \times \beta_b(\boldsymbol{\gamma_b b_i})\big) + \boldsymbol{\delta_b}(\boldsymbol{\gamma_b}^{\boldsymbol{T}} \circ \boldsymbol{b_i})\right)$ |
| PF | $expit(\boldsymbol{\gamma_f f_i})$ | $expit(\boldsymbol{\gamma_f f_i} + \beta_f)$ |
| BOTH+PF | $expit(\boldsymbol{\gamma_b b_i} + \boldsymbol{\gamma_f f_i})$ | $expit\left(\boldsymbol{\gamma_f f_i} + logit\big(expit(\boldsymbol{\gamma_b b_i}) \times \beta_b(\boldsymbol{\gamma_b b_i})\big) + \boldsymbol{\delta_b}(\boldsymbol{\gamma_b}^{\boldsymbol{T}} \circ \boldsymbol{b_i})\right)$ |
| BOTH+PF+M | $expit(\boldsymbol{\gamma_b b_i} + \boldsymbol{\gamma_f f_i})$ | $expit\big(\boldsymbol{\gamma_f f_i} + logit\big(expit(\boldsymbol{\gamma_b b_i}) \times \beta_b(\boldsymbol{\gamma_b b_i})\big) + \boldsymbol{\delta_b}(\boldsymbol{\gamma_b}^{\boldsymbol{T}} \circ \boldsymbol{b_i}) + \boldsymbol{\delta_m}^{\boldsymbol{T}}\boldsymbol{m_i}\big)$ |

*Table 3: Coefficients used to generate the data in each scenario. Prognostic factors with coefficients of 0.5 and -0.5 are dichotomized.*

| Scenario | $\gamma_b$ | $\gamma_f$ | $\beta_f$ | $\beta_b(\boldsymbol{\gamma_b b_i})$ | $\delta_b$ | $\delta_m$ |
|---|---|---|---|---|---|---|
| BOTH a) | $(1,0.9,0.7,0.1,-0.5,0.5)^T$ | - | - | 0.5 | $(-0.1,-0.05,0,0,0.05,0.1)^T$ | |
| BOTH b) | $(1,0.9,0.7,0.1,-0.5,0.5)^T$ | - | - | 0.5 | $(-0.1,-0.05,0,0,0.05,0.1)^T$ | |
| BOTH c) | $(1,0.9,0.7,0.1,-0.5,0.5)^T$ | - | - | 0.5 | $(-0.1,-0.05,0,0,0.05,0.1)^T$ | |
| BOTH d) | $(1,0.9,0.7,0.1,-0.5,0.5)^T$ | - | - | $expit(\gamma_b b_i)^{7/2}$ | $(-0.1,-0.05,0,0,0.05,0.1)^T$ | |
| BOTH e) | $(1,0.9,0.7,0.1,-0.5,0.5)^T$ | - | - | $1-2/3\times sqrt(expit(\gamma_b b_i))$ | $(-0.1,-0.05,0,0,0.05,0.1)^T$ | |
| BOTH f) | $(1,0.9,0.7,0.1,-0.5,0.5)^T$ | - | - | 0.5 | $(-0.25,-0.15,-0.05,0.05,0.15,0.25)^T$ | |
| BOTH g) | $(1,0.9,0.7,0.1,-0.5,0.5)^T$ | - | - | 0.5 | $(-0.1,-0.05,0,0,0.05,0.1)^T$ | |
| PF a) | - | $(1,0.9,0.7,0.1,-0.5,0.5)^T$ | log(0.5) | - | - | |
| PF b) | - | $(1,0.9,0.7,0.1,-0.5,0.5)^T$ | log(0.5) | - | - | |
| PF c) | - | $(1,0.9,0.7,0.1,-0.5,0.5)^T$ | log(0.5) | - | - | |
| PF d) | - | $(1,0.9,0.7,0.1,-0.5,0.5)^T$ | log(0.5) | - | - | |
| BOTHPF a) | $(1,0.9,0.1,-0.5)^T$ | $(0.7,0.5)^T$ | - | 0.5 | $(-0.1,-0.05,0.05,0.1)^T$ | |
| BOTHPF b) | $(1,0.9,0.1,-0.5)^T$ | $(0.7,0.5)^T$ | - | 0.5 | $(-0.1,-0.05,0.05,0.1)^T$ | |
| BOTHPF c) | $(1,0.9,0.1,-0.5)^T$ | $(0.7,0.5)^T$ | - | 0.5 | $(-0.1,-0.05,0.05,0.1)^T$ | |
| BOTHPF d) | $(1,0.9,0.1,-0.5)^T$ | $(0.7,0.5)^T$ | - | $expit(\boldsymbol{\gamma_b b_i})^{7/2}$ | $(-0.1,-0.05,0.05,0.1)^T$ | |
| BOTHPF e) | $(1,0.9,0.1,-0.5)^T$ | $(0.7,0.5)^T$ | - | $1-2/3\times sqrt(expit(\boldsymbol{\gamma_b b_i}))$ | $(-0.1,-0.05,0.05,0.1)^T$ | |
| BOTHPF f) | $(1,0.9,0.1,-0.5)^T$ | $(0.7,0.5)^T$ | - | 0.5 | $(-0.25,-0.1,0.1,0.25)^T$ | |
| BOTHPF g) | $(1,0.9,0.1,-0.5)^T$ | $(0.7,0.5)^T$ | - | 0.5 | $(-0.1,-0.05,0.05,0.1)^T$ | |
| BOTHPFM a) | $(1,0.9,-0.5)^T$ | $(0.7,0.5)^T$ | - | 0.5 | $(-0.1,0.05,0.1)^T$ | log(0.75) |
| BOTHPFM b) | $(1,0.9,-0.5)^T$ | $(0.7,0.5)^T$ | - | 0.5 | $(-0.1,0.05,0.1)^T$ | log(0.75 |
| BOTHPFM c) | $(1,0.9,-0.5)^T$ | $(0.7,0.5)^T$ | - | 0.5 | $(-0.1,0.05,0.1)^T$ | log(0.75 |
| BOTHPFM d) | $(1,0.9,-0.5)^T$ | $(0.7,0.5)^T$ | - | $expit(\boldsymbol{\gamma_b b_i})^{7/2}$ | $(-0.1,0.05,0.1)^T$ | log(0.75 |
| BOTHPFM e) | $(1,0.9,-0.5)^T$ | $(0.7,0.5)^T$ | - | $1-2/3\times sqrt(expit(\boldsymbol{\gamma_b b_i}))$ | $(-0.1,0.05,0.1)^T$ | log(0.75 |
| BOTHPFM f) | $(1,0.9,-0.5)^T$ | $(0.7,0.5)^T$ | - | 0.5 | $(-0.25,0.1,0.25)^T$ | log(0.75 |
| BOTHPFM g) | $(1,0.9,-0.5)^T$ | $(0.7,0.5)^T$ | - | 0.5 | $(-0.1,0.05,0.1)^T$ | log(0.75 |

### 7.3 Tables of results

Here we show tables with results corresponding to figures 6 and 7 in the main text.

*Table 4: RMSEs of the predicted treatment effect as a risk difference of each analysis method for all the scenarios. The first number in each cell is the mean RMSE of 1000 simulation runs, the numbers in the brackets are the 5th to 95th percentile intervals of the 1000 simulation runs. For each sample size and scenario, the better performing method in terms of mean RMSE is shown with bold numbers*

| Scenario | N = 500 | | N = 2000 | | N = 5000 | |
|---|---|---|---|---|---|---|
| | EM | RM | EM | RM | EM | RM |
| BOTH a) | 0.09 (0.06, 0.13) | **0.07 (0.04, 0.1)** | 0.06 (0.04, 0.08) | **0.05 (0.03, 0.06)** | 0.04 (0.03, 0.05) | 0.04 (0.03, 0.05) |
| BOTH b) | 0.1 (0.07, 0.14) | **0.08 (0.06, 0.11)** | 0.07 (0.05, 0.09) | **0.06 (0.05, 0.08)** | 0.06 (0.05, 0.07) | 0.06 (0.05, 0.06) |
| BOTH c) | 0.07 (0.04, 0.1) | **0.05 (0.03, 0.09)** | 0.04 (0.03, 0.06) | **0.03 (0.02, 0.05)** | 0.03 (0.02, 0.04) | **0.02 (0.02, 0.03)** |
| BOTH d) | 0.08 (0.05, 0.12) | **0.07 (0.04, 0.1)** | 0.05 (0.03, 0.06) | **0.04 (0.03, 0.05)** | 0.04 (0.03, 0.05) | **0.03 (0.02, 0.04)** |
| BOTH e) | 0.1 (0.06, 0.14) | **0.08 (0.05, 0.11)** | 0.06 (0.04, 0.08) | **0.05 (0.04, 0.07)** | 0.05 (0.04, 0.06) | **0.04 (0.04, 0.05)** |
| BOTH f) | 0.09 (0.06, 0.13) | **0.08 (0.05, 0.11)** | 0.06 (0.04, 0.08) | **0.05 (0.04, 0.07)** | **0.04 (0.03, 0.05)** | 0.05 (0.04, 0.06) |
| BOTH g) | 0.1 (0.07, 0.14) | **0.09 (0.07, 0.12)** | 0.07 (0.06, 0.09) | 0.07 (0.06, 0.08) | 0.07 (0.06, 0.07) | **0.06 (0.06, 0.07)** |
| PF a) | 0.08 (0.03, 0.16) | **0.05 (0.02, 0.1)** | 0.04 (0.02, 0.09) | **0.03 (0.01, 0.05)** | 0.03 (0.01, 0.06) | **0.02 (0.01, 0.03)** |
| PF b) | 0.07 (0.03, 0.12) | **0.05 (0.02, 0.09)** | 0.04 (0.02, 0.06) | **0.02 (0.01, 0.04)** | 0.03 (0.01, 0.04) | **0.02 (0.01, 0.03)** |
| PF c) | 0.06 (0.02, 0.1) | **0.04 (0.02, 0.09)** | 0.03 (0.01, 0.06) | **0.02 (0.01, 0.04)** | 0.02 (0.01, 0.04) | 0.02 (0.01, 0.03) |
| PF d) | 0.09 (0.03, 0.17) | **0.06 (0.03, 0.1)** | 0.05 (0.03, 0.09) | **0.04 (0.03, 0.06)** | 0.04 (0.03, 0.06) | **0.03 (0.02, 0.04)** |
| BOTHPF a) | 0.08 (0.05, 0.13) | **0.06 (0.04, 0.1)** | 0.05 (0.03, 0.07) | **0.04 (0.03, 0.06)** | 0.04 (0.02, 0.05) | 0.04 (0.03, 0.05) |
| BOTHPF b) | 0.09 (0.06, 0.14) | 0.09 (0.07, 0.12) | **0.06 (0.04, 0.08)** | 0.08 (0.07, 0.08) | **0.04 (0.04, 0.05)** | 0.07 (0.07, 0.08) |
| BOTHPF c) | 0.07 (0.04, 0.1) | **0.05 (0.03, 0.09)** | 0.04 (0.02, 0.06) | **0.03 (0.02, 0.05)** | 0.03 (0.02, 0.04) | 0.03 (0.02, 0.04) |
| BOTHPF d) | 0.08 (0.05, 0.12) | **0.07 (0.05, 0.11)** | 0.05 (0.03, 0.06) | **0.04 (0.03, 0.06)** | **0.03 (0.02, 0.04)** | 0.04 (0.03, 0.04) |
| BOTHPF e) | 0.09 (0.06, 0.13) | **0.08 (0.06, 0.11)** | 0.06 (0.04, 0.08) | 0.06 (0.05, 0.08) | **0.04 (0.03, 0.05)** | 0.06 (0.05, 0.07) |
| BOTHPF f) | 0.09 (0.05, 0.13) | **0.07 (0.05, 0.11)** | 0.05 (0.03, 0.07) | 0.05 (0.05, 0.07) | **0.03 (0.02, 0.05)** | 0.05 (0.04, 0.06) |
| BOTHPF g) | 0.09 (0.06, 0.13) | **0.08 (0.06, 0.11)** | 0.07 (0.05, 0.08) | **0.06 (0.05, 0.07)** | 0.06 (0.05, 0.07) | 0.06 (0.05, 0.06) |
| BOTHPFM a) | 0.09 (0.05, 0.13) | 0.09 (0.07, 0.11) | **0.05 (0.03, 0.07)** | 0.07 (0.06, 0.08) | **0.03 (0.02, 0.05)** | 0.07 (0.06, 0.07) |
| BOTHPFM b) | 0.1 (0.06, 0.13) | 0.1 (0.09, 0.13) | **0.06 (0.04, 0.08)** | 0.09 (0.09, 0.1) | **0.04 (0.04, 0.06)** | 0.09 (0.09, 0.1) |
| BOTHPFM c) | 0.07 (0.04, 0.11) | **0.06 (0.04, 0.1)** | 0.04 (0.03, 0.06) | 0.04 (0.03, 0.06) | **0.03 (0.02, 0.04)** | 0.04 (0.03, 0.05) |
| BOTHPFM d) | 0.08 (0.05, 0.12) | 0.08 (0.05, 0.11) | 0.05 (0.03, 0.06) | 0.05 (0.04, 0.07) | **0.03 (0.02, 0.04)** | 0.04 (0.04, 0.05) |
| BOTHPFM e) | 0.1 (0.06, 0.14) | 0.1 (0.08, 0.13) | **0.05 (0.03, 0.08)** | 0.09 (0.08, 0.1) | **0.04 (0.03, 0.05)** | 0.09 (0.08, 0.09) |

| BOTHPFM f) | 0.09 (0.05, 0.13) | 0.09 (0.07, 0.12) | **0.05 (0.03, 0.07)** | 0.08 (0.07, 0.09) | **0.03 (0.02, 0.05)** | 0.07 (0.07, 0.08) |
|---|---|---|---|---|---|---|
| BOTHPFM g) | 0.1 (0.07, 0.14) | 0.1 (0.08, 0.13) | **0.07 (0.06, 0.09)** | 0.08 (0.08, 0.09) | **0.06 (0.06, 0.07)** | 0.08 (0.08, 0.09) |

*Table 5: Bias of the predicted treatment effect as a risk difference of each analysis method for all the scenarios. The first number in each cell is the mean bias of 1000 simulation runs, the numbers in the brackets are the 5th to 95th percentile intervals of the 1000 simulation runs. For each sample size and scenario, the better performing method in terms of mean bias is shown with bold numbers.*

| Scenario | N = 500 | | N = 2000 | | N = 5000 | |
|---|---|---|---|---|---|---|
| | EM | RM | EM | RM | EM | RM |
| BOTH a) | -0.02 (-0.08, 0.05) | **0 (-0.06, 0.06)** | 0 (-0.03, 0.03) | 0 (-0.03, 0.03) | 0 (-0.02, 0.02) | 0 (-0.02, 0.02) |
| BOTH b) | -0.01 (-0.07, 0.05) | **0 (-0.06, 0.06)** | 0 (-0.03, 0.03) | 0 (-0.03, 0.03) | 0 (-0.02, 0.02) | 0 (-0.02, 0.02) |
| BOTH c) | -0.01 (-0.05, 0.03) | **0 (-0.04, 0.04)** | -0.01 (-0.03, 0.01) | **0 (-0.02, 0.02)** | 0 (-0.01, 0.01) | 0 (-0.01, 0.01) |
| BOTH d) | -0.01 (-0.07, 0.04) | **0 (-0.06, 0.05)** | 0 (-0.03, 0.02) | 0 (-0.03, 0.03) | 0 (-0.02, 0.02) | 0 (-0.02, 0.02) |
| BOTH e) | -0.01 (-0.08, 0.05) | **0 (-0.06, 0.06)** | 0 (-0.04, 0.03) | 0 (-0.03, 0.03) | 0 (-0.02, 0.02) | 0 (-0.02, 0.02) |
| BOTH f) | -0.01 (-0.08, 0.05) | **0 (-0.06, 0.06)** | 0 (-0.04, 0.03) | 0 (-0.03, 0.03) | 0 (-0.02, 0.02) | 0 (-0.02, 0.02) |
| BOTH g) | -0.01 (-0.08, 0.05) | **0 (-0.06, 0.06)** | 0 (-0.03, 0.03) | 0 (-0.03, 0.03) | 0 (-0.02, 0.02) | 0 (-0.02, 0.02) |
| PF a) | -0.03 (-0.12, 0.07) | **0 (-0.07, 0.06)** | -0.01 (-0.06, 0.04) | **0 (-0.03, 0.04)** | -0.01 (-0.04, 0.03) | **0 (-0.02, 0.02)** |
| PF b) | -0.02 (-0.08, 0.04) | **0 (0.06, 0.06)** | -0.01 (-0.03, 0.02) | **0 (-0.03, 0.03)** | 0 (-0.02, 0.02) | 0 (-0.02, 0.02) |
| PF c) | -0.01 (-0.05, 0.03) | **0 (-0.04, 0.04)** | -0.01 (-0.03, 0.02) | **0 (-0.02, 0.02)** | 0 (-0.02, 0.01) | 0 (-0.01, 0.01) |
| PF d) | -0.03 (-0.12, 0.08) | **0 (-0.07, 0.07)** | -0.01 (-0.06, 0.04) | **0 (-0.03, 0.04)** | -0.01 (-0.04, 0.03) | **0 (-0.02, 0.02)** |
| BOTHPF a) | -0.02 (-0.08, 0.05) | **0 (-0.06, 0.06)** | -0.01 (-0.04, 0.03) | **0 (-0.03, 0.03)** | 0 (-0.02, 0.02) | 0 (-0.02, 0.02) |
| BOTHPF b) | -0.01 (-0.07, 0.04) | **0 (-0.06, 0.06)** | 0 (-0.03, 0.03) | 0 (-0.03, 0.03) | 0 (-0.02, 0.02) | 0 (-0.02, 0.02) |
| BOTHPF c) | -0.01 (-0.05, 0.03) | **0 (-0.04, 0.04)** | -0.01 (-0.02, 0.01) | **0 (-0.02, 0.02)** | 0 (-0.01, 0.01) | 0 (-0.01, 0.01) |
| BOTHPF d) | -0.01 (-0.06, 0.04) | **0 (-0.05, 0.05)** | 0 (-0.03, 0.02) | 0 (-0.03, 0.02) | 0 (-0.02, 0.01) | 0 (-0.02, 0.02) |
| BOTHPF e) | -0.01 (-0.08, 0.05) | **0 (-0.06, 0.06)** | 0 (-0.04, 0.03) | 0 (-0.03, 0.03) | 0 (-0.02, 0.02) | 0 (-0.02, 0.02) |
| BOTHPF f) | -0.02 (-0.08, 0.05) | **0 (-0.06, 0.06)** | -0.01 (-0.04, 0.03) | **0 (-0.03, 0.03)** | 0 (-0.02, 0.02) | 0 (-0.02, 0.02) |
| BOTHPF g) | -0.02 (-0.08, 0.05) | **0 (-0.07, 0.06)** | -0.01 (-0.04, 0.03) | **0 (-0.03, 0.03)** | 0 (-0.02, 0.02) | 0 (-0.02, 0.02) |
| BOTHPFM a) | -0.02 (-0.08, 0.05) | **0 (-0.07, 0.06)** | -0.01 (-0.04, 0.02) | **0 (-0.03, 0.03)** | 0 (-0.02, 0.02) | 0 (-0.02, 0.02) |
| BOTHPFM b) | -0.01 (-0.07, 0.05) | **0 (-0.06, 0.06)** | 0 (-0.03, 0.03) | 0 (-0.03, 0.03) | 0 (-0.02, 0.02) | 0 (-0.02, 0.02) |
| BOTHPFM c) | -0.01 (-0.05, 0.03) | **0 (-0.04, 0.04)** | 0 (-0.02, 0.02) | 0 (-0.02, 0.02) | 0 (-0.01, 0.01) | 0 (-0.01, 0.01) |
| BOTHPFM d) | -0.01 (-0.07, 0.04) | **0 (-0.06, 0.05)** | 0 (-0.03, 0.02) | 0 (-0.03, 0.03) | 0 (-0.02, 0.02) | 0 (-0.02, 0.02) |
| BOTHPFM e) | -0.01 (-0.08, 0.05) | **0 (-0.07, 0.07)** | 0 (-0.03, 0.03) | 0 (-0.03, 0.03) | 0 (-0.02, 0.02) | 0 (-0.02, 0.02) |
| BOTHPFM f) | -0.02 (-0.08, 0.05) | **0 (-0.06, 0.06)** | -0.01 (-0.04, 0.02) | **0 (-0.03, 0.03)** | 0 (-0.02, 0.02) | 0 (-0.02, 0.02) |

| BOTHPFM g) | -0.02 (-0.08, 0.05) | **0 (-0.07, 0.06)** | -0.01 (-0.04, 0.02) | **0 (-0.03, 0.03)** | 0 (-0.02, 0.02) | 0 (-0.02, 0.02) |
|---|---|---|---|---|---|---|

### 7.4 Explanation of figure 7

Figure 7 in the main text presents the mean predicted RDs against the true simulated baseline risks. Prediction errors of the logistic regression model vary with the baseline risk, with predictions being more inaccurate for participant at high or low baseline risks rather than those in the middle of the range. Simulated baseline risk for a participant depends on their characteristics and there is not a one-to-one relationship between the predicted treatment effects and baseline risk. This is why we present the mean predicted RD, as estimated from a kernel regression across participants with the same baseline risk but different combination of values of their characteristics. Per scenario we get 1000 mean predicted RD, one from each simulation run. We present the area between the $5^{th}$ and $95^{th}$ percentile of the mean predicted RD as a function of the baseline risk to show where typical predicted treatment effects lie.

### 7.5 Bias for high-risk participants

We calculated the bias of the predicted RD in our simulation study for each approach for the quartile of participants with the highest baseline risk. Results are shown in figure 8 and in table 6.

*Figure 8: violin plots of the bias of the predicted treatment effect as a risk difference for the highest quartile of baseline risk participants of each analysis method for all the scenarios.*

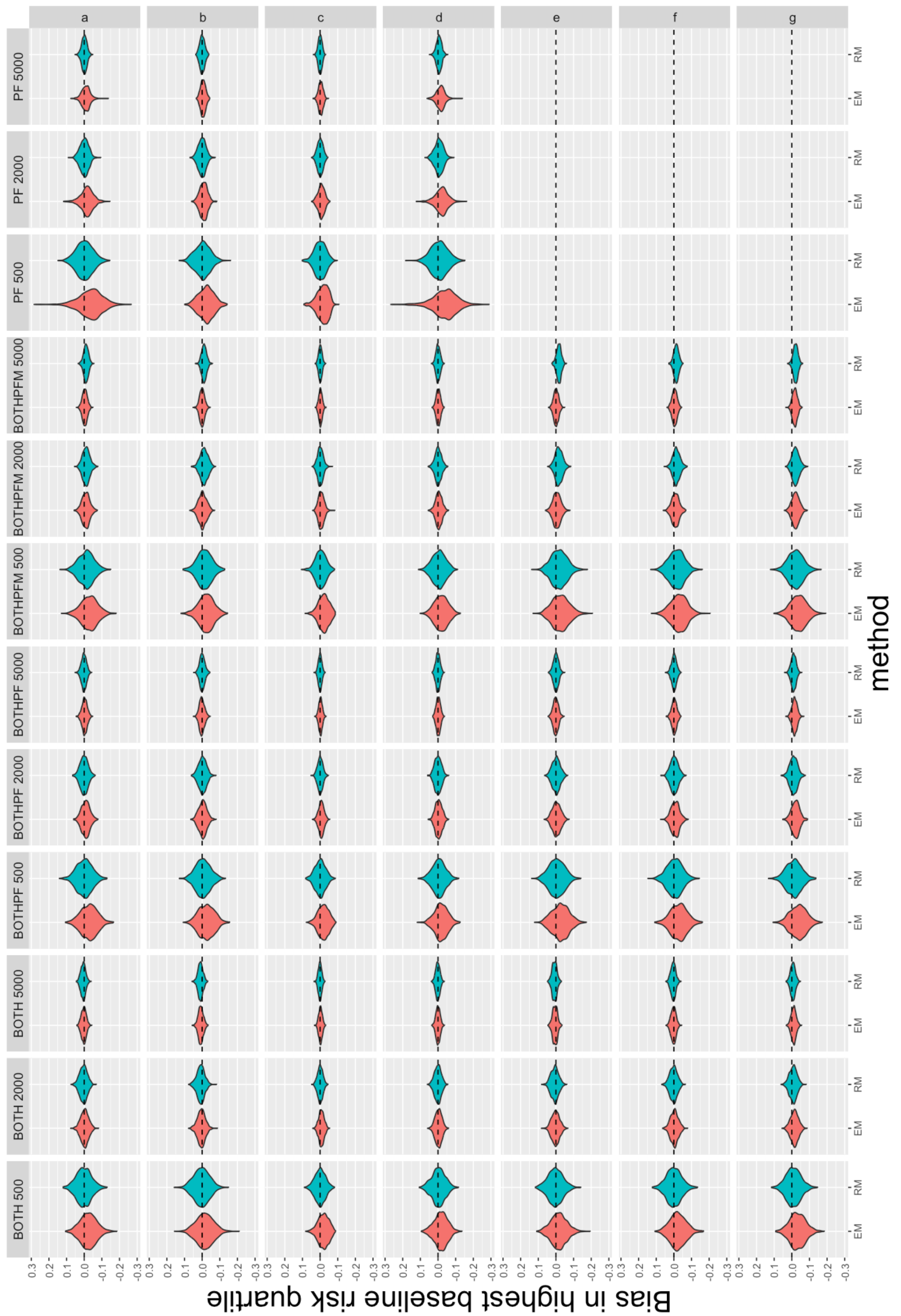

*Table 6: Bias of the predicted treatment effect as a risk difference for the highest quartile of baseline risk participants of each analysis method for all the scenarios. The first number in each cell is the mean bias of 1000 simulation runs, the numbers in the brackets are the 5th to 95th percentile intervals of the 1000 simulation runs. For each sample size and scenario, the better performing method in terms of mean bias is shown with bold numbers.*

| Scenario | N = 500 | | N = 2000 | | N = 5000 | |
|---|---|---|---|---|---|---|
| | EM | RM | EM | RM | EM | RM |
| BOTH a) | -0.02 (-0.1, 0.05) | **0 (-0.07, 0.07)** | **0 (-0.04, 0.03)** | 0.01 (-0.03, 0.04) | **0 (-0.02, 0.03)** | 0.01 (-0.02, 0.03) |
| BOTH b) | -0.02 (-0.1, 0.07) | **0 (-0.07, 0.06)** | 0.01 (-0.03, 0.04) | 0.01 (-0.03, 0.05) | -0.01 (-0.02, 0.03) | 0.01 (-0.01, 0.03) |
| BOTH c) | -0.02 (-0.07, 0.03) | **0 (-0.05, 0.05)** | -0.01 (-0.03, 0.02) | **0 (-0.03, 0.02)** | 0 (-0.02, 0.01) | 0 (-0.02, 0.02) |
| BOTH d) | -0.02 (-0.08, 0.05) | **0 (-0.06, 0.06)** | 0 (-0.03, 0.03) | 0 (-0.03, 0.03) | 0 (-0.02, 0.02) | 0 (-0.02, 0.02) |
| BOTH e) | -0.02 (-0.1, 0.06) | **0 (-0.07, 0.07)** | **0 (-0.03, 0.04)** | 0.01 (-0.03, 0.04) | 0.01 (-0.02, 0.03) | 0.01 (-0.01, 0.03) |
| BOTH f) | -0.02 (-0.1, 0.05) | **0 (-0.07, 0.07)** | 0 (-0.04, 0.03) | 0 (-0.03, 0.04) | 0 (-0.02, 0.03) | 0 (-0.02, 0.03) |
| BOTH g) | -0.03 (-0.11, 0.04) | **-0.01 (-0.08, 0.06)** | -0.02 (-0.05, 0.02) | **-0.01 (-0.04, 0.03)** | -0.01 (-0.03, 0.01) | -0.01 (-0.03, 0.02) |
| PF a) | -0.04 (-0.14, 0.08) | **0 (-0.08, 0.08)** | -0.02 (-0.07, 0.05) | **0 (-0.04, 0.04)** | -0.01 (-0.05, 0.03) | **0 (-0.02, 0.02)** |
| PF b) | -0.02 (-0.1, 0.05) | **0 (-0.07, 0.07)** | -0.01 (-0.04, 0.03) | **0 (-0.04, 0.04)** | -0.01 (-0.03, 0.02) | **0 (-0.02, 0.02)** |
| PF c) | -0.02 (-0.06, 0.04) | **0 (-0.05, 0.05)** | -0.01 (-0.04, 0.03) | **0 (-0.04, 0.03)** | 0 (-0.02, 0.02) | 0 (-0.02, 0.02) |
| PF d) | -0.04 (-0.14, 0.08) | **-0.01 (-0.09, 0.07)** | -0.02 (-0.08, 0.04) | **-0.01 (-0.02, 0.02)** | -0.01 (-0.05, 0.03) | -0.01 (-0.03, 0.02) |
| BOTHPF a) | -0.03 (-0.1, 0.05) | **0 (-0.07, 0.07)** | -0.01 (-0.05, 0.03) | **0 (-0.03, 0.03)** | 0 (-0.03, 0.02) | 0 (-0.02, 0.03) |
| BOTHPF b) | -0.03 (-0.1, 0.05) | **-0.01 (-0.08, 0.07)** | 0 (-0.04, 0.03) | 0 (-0.04, 0.04) | 0 (-0.02, 0.02) | 0 (-0.03, 0.02) |
| BOTHPF c) | -0.02 (-0.07, 0.03) | **0 (-0.05, 0.05)** | -0.01 (-0.03, 0.02) | **0 (-0.02, 0.02)** | 0 (-0.02, 0.02) | 0 (-0.02, 0.02) |
| BOTHPF d) | -0.02 (-0.08, 0.05) | **-0.01 (-0.07, 0.06)** | 0 (-0.03, 0.03) | 0 (-0.03, 0.03) | 0 (-0.02, 0.02) | 0 (-0.02, 0.02) |
| BOTHPF e) | -0.03 (-0.11, 0.05) | **-0.01 (-0.08, 0.07)** | 0 (-0.04, 0.03) | 0 (-0.04, 0.04) | 0 (-0.02, 0.03) | 0 (-0.02, 0.02) |
| BOTHPF f) | -0.03 (-0.1, 0.05) | **-0.01 (-0.08, 0.07)** | -0.01 (-0.05, 0.03) | **0 (-0.04, 0.03)** | 0 (-0.03, 0.02) | 0 (-0.02, 0.02) |
| BOTHPF g) | -0.04 (-0.11, 0.04) | **-0.01 (-0.08, 0.06)** | -0.02 (-0.05, 0.02) | **-0.01 (-0.04, 0.02)** | -0.02 (-0.04, 0.01) | **-0.01 (-0.03, 0.01)** |
| BOTHPFM a) | -0.03 (-0.11, 0.05) | **-0.01 (-0.08, 0.06)** | -0.01 (-0.05, 0.03) | -0.01 (-0.05, 0.02) | **0 (-0.03, 0.02)** | -0.01 (-0.03, 0.01) |
| BOTHPFM b) | -0.02 (-0.1, 0.05) | **0 (-0.09, 0.06)** | **0 (-0.04, 0.03)** | -0.01 (-0.05, 0.03) | **0 (-0.02, 0.02)** | -0.01 (-0.03, 0.01) |
| BOTHPFM c) | -0.02 (-0.07, 0.03) | **0 (-0.05, 0.05)** | -0.01 (-0.03, 0.02) | **0 (-0.03, 0.02)** | 0 (-0.02, 0.01) | 0 (-0.02, 0.01) |
| BOTHPFM d) | -0.02 (-0.08, 0.04) | **-0.01 (-0.07, 0.06)** | 0 (-0.04, 0.03) | 0 (-0.03, 0.03) | 0 (-0.02, 0.02) | 0 (-0.02, 0.02) |
| BOTHPFM e) | -0.03 (0.11, 0.05) | **-0.02 (-0.09, 0.06)** | **-0.01 (-0.04, 0.03)** | -0.02 (-0.05, 0.02) | **0 (-0.02, 0.02)** | -0.02 (-0.04, 0.01) |
| BOTHPFM f) | -0.03 (-0.11, 0.05) | **-0.02 (-0.09, 0.06)** | -0.01 (-0.05, 0.03) | -0.01 (-0.05, 0.02) | **0 (-0.03, 0.02)** | -0.01 (-0.04, 0.01) |
| BOTHPFM g) | -0.04 (-0.11, 0.04) | **-0.02 (-0.09, 0.05)** | -0.02 (-0.06, 0.01) | -0.02 (-0.06, 0.01) | -0.02 (-0.04, 0.01) | -0.02 (-0.03, 0) |